\documentclass[epj]{svjour}

\usepackage{amssymb}
\usepackage{amsmath}
\usepackage{graphicx}

\begin{document}

\title{Modified Chaplygin Gas as a Unified Dark Matter and Dark Energy Model and Cosmic Constraints}

\author{Lixin Xu\inst{1,2,3} \thanks{lxxu@dlut.edu.cn} \and Yuting Wang\inst{1,4} \and Hyerim Noh \inst{3}
}

\institute{Institute of Theoretical Physics, Dalian University of Technology, Dalian,
116024, People's Republic of China \and College of Advanced Science \& Technology, 
Dalian University of Technology, Dalian, 116024, People's Republic of China \and Korea Astronomy and Space Science Institute,
Yuseong Daedeokdaero 776,
Daejeon 305-348,
R. Korea  \and Institute of Cosmology \& Gravitation,
University of Portsmouth, Portsmouth, PO1 3FX, United Kingdom}

\abstract{
A modified Chaplygin gas model (MCG), $\rho_{MCG}/\rho_{MCG0}=[B_{s}+(1-B_{s})a^{-3(1+B)(1+\alpha)}]^{1/(1+\alpha)}$, as a unified dark matter model and dark energy model is constrained by using current available cosmic observational data points which include type Ia supernovae, baryon acoustic oscillation and the seventh year full WMAP data points. As a contrast to the consideration in the literatures, we {\it do not} separate the MCG into two components, i.e. dark mater and dark energy component, but we take it as a whole energy component-a unified dark sector. By using  Markov Chain Monte Carlo method, a tight constraint is obtained: $\alpha= 0.000727_{-    0.00140-    0.00234}^{+    0.00142+    0.00391}$,
$B=0.000777_{-    0.000302-    0.000697}^{+    0.000201+    0.000915}$ and
$B_s= 0.782_{-    0.0162-    0.0329}^{+    0.0163+    0.0307}$ .}

\maketitle

\section{Introduction}

Under the assumption of Einstein gravity theory, from the observed energy components $T^{obs}_{\mu\nu}$ and the geometric structure $G_{\mu\nu}$, the un-observed energy component is defined as a dark fluid
\begin{equation}
T^{dark}_{\mu\nu}=\frac{1}{8\pi G}G_{\mu\nu}-T^{obs}_{\mu\nu}.\label{eq:dark}
\end{equation} 
 In the literatures, a dark fluid is usually separated into dark matter and dark energy parts, for recent reviews on dark energy please see \cite{ref:DEReview1,ref:DEReview2,ref:DEReview3,ref:DEReview4,ref:DEReview5,ref:DEReview6,ref:DEReview7}. For gravitational attraction, the dark matter component is responsible for galaxy structure formation, and dark energy provides repulsive force for current accelerated expansion \cite{ref:Riess98,ref:Perlmuter99}. However, we can not observe the dark sectors directly in cosmology. From the above equation (\ref{eq:dark}), one can easily see the degeneracies between dark fluid and gravity theory and can understand the reasons of why a kind of modified gravity theory, for example $f(R)$ gravity theory, can realize current accelerated expansion. But what is the content of the dark fluid? Be honest, one can give a long list of candidates which include dark matter and dark energy models, but the last word is still empty until now. If the Higgs particles were founded in LHC, the scalar field, quintessence, would be a competitive candidate. But another possibility is that the dark fluid is a mixture of dark matter and dark energy components, or it is just one exotic unknown fluid. This property, dubbed dark degeneracy, has been discussed by the authors \cite{ref:darkdeneracy}. A virtue of the unified dark fluid model is that the so-called coincidence problem is removed completely.  

A modified Chaplygin gas (MCG) model \cite{ref:MCG}, which is a unified dark matter and dark energy model, is an example of dark degeneracy. MCG model is a variant of Generalized Chaplygin gas (GCG) \cite{ref:GCG} which is a generalization of Chaplygin gas (CG) \cite{ref:CG}. MCG has been discussed in many perspectives extensively \cite{ref:MCGall,ref:Lu,ref:LuMCG}.

In our previous work \cite{ref:Lu}, the energy density of MCG is decomposed into two components: one part, which evolves with the calling relation $a^{-3}$, is dark matter; and the remaining part is dark energy. With this decomposition, by using $182$ Gold SN Ia, 3-year WMAP and SDSS BAO, the best fit values of the model parameters were obtained: $B_s=0.822$, $\alpha=1.724$ and $B=-0.085$. In our work \cite{ref:LuMCG}, we used CMB shift parameters, BAO, SN Ia Union 2, observational Hubble data and cluster X-ray gas mass fraction (CBF) to constrain the model space: $B_{s}=0.7788^{+0.0736}_{-0.0723}$ ($1\sigma$) $^{+0.0918}_{-0.0904}$ $(2\sigma)$, $\alpha=0.1079^{+0.3397}_{-0.2539}$ ($1\sigma$) $^{+0.4678}_{-0.2911}$ $(2\sigma)$, $B=0.00189^{+0.00583}_{-0.00756}$ ($1\sigma$) $^{+0.00660}_{-0.00915}$ $(2\sigma)$, where MCG was taken as a unified dark sector. But in the CBF, BAO and CMB constraints, an effective matter density $\Omega_{m}=\Omega_{b}+(1-\Omega_{b}-\Omega_{r}-\Omega_{k})(1-B_{s})^{1/(1+\alpha)}$ was used. So in some senses, the potential decomposition was employed too. When one just considers the background evolution and doesn't need any kind of definition of $\Omega_m$, one doesn't worry about this issue. But one has to take into account decomposition carefully, when the perturbations of energy component is involved. The problem comes from the fact that the decomposition is not unique. One can give many kinds of decomposition, because of the lack of physical principle to do a decomposition. The worst is that the evolutions of perturbations strongly depend on the decomposition and definition of dark matter. As a contrast, in this paper, we shall {\it not} take any decomposition like that in our previous work \cite{ref:Lu,ref:LuMCG}, i.e. we take MCG as a whole energy component entirely. Furthermore, the perturbations of MCG will also be included. In our previous works \cite{ref:Lu,ref:LuMCG}, the background evolution information and the shift parameters $R$ and $l_a$ and $z_\ast$ but not the full information from CMB was used. One can expect that, by combining the full information from CMB, a tighter constraint would be obtained. So, in this paper, we shall constrain the background evolution by using BAO and SN Ia data points, and its perturbation evolution by using the full CMB data. Finally, the model parameter space is obtained.

This paper is structured as follows. In section \ref{sec:MCG}, the equation of state (EoS) and adiabatic sound speed of MCG are shown. Meanwhile, the Friedmann equation and perturbation equation are given. In section \ref{sec:method}, the constraint method and results are presented. A summary is given in Section \ref{ref:conclusion}.

\section{Main equations in modified Chaplygin gas model}   \label{sec:MCG}  

The MCG is characterized by its equation of state (EoS)
\begin{equation}
p_{MCG}=B\rho_{MCG}-A/\rho^{\alpha}_{MCG}
\end{equation} 
where $B$, $A$ and $\alpha$ are model parameters. It is obvious that GCG is recovered when the value of $B$ is zero. And, a cosmological constant $\Lambda$ is reduced when $\alpha=-1$ and $A=1+B$. Also, if $A=0$ is respected, MCG looks like a perfect fluid with EoS $w=B$, for example a quintessence model. However, if MCG as a unified dark sector, it would not happen. Because it would look like a combination of cold dark matter and a simple cosmological constant. In general, for a spatially non-flat FRW universe, the metric is written as
\begin{equation}
ds^{2}=-dt^{2}+a^{2}(t)\left[\frac{1}{1-kr^{2}}dr^{2}+r^{2}(d\theta^{2}+\sin^{2}\theta d\phi^{2})\right],
\end{equation}
where $k=0,\pm 1$ is the three-dimensional curvature and $a$ is the scale factor. Considering the energy conservation of MCG, one can rewrite the MCG energy density as
\begin{equation}
\rho_{MCG}=\rho_{MCG0}\left[B_{s}+(1-B_{s})a^{-3(1+B)(1+\alpha)}\right]^{\frac{1}{1+\alpha}}\label{eq:mcg}
\end{equation}
for $B\neq -1$, where $B_{s}=A/(1+B)\rho^{1+\alpha}_{MCG0}$. We shall take $B_s$, $B$ and $\alpha$ as MCG model parameters in this paper. Form Eq. (\ref{eq:mcg}), one can find that $0\le B_s \le 1$ is demanded to keep the positivity of energy density. If $\alpha=0$ and $B=0$ in Eq. (\ref{eq:mcg}), the standard $\Lambda$CDM model is recovered. Taking MCG as a unified component, one has the Friedmann equation
\begin{eqnarray}
H^{2}&=&H^{2}_{0}\left\{\Omega_{b}a^{-3}+\Omega_{r}a^{-4}+\Omega_{k}a^{-2}\right.\nonumber\\
&+&\left.(1-\Omega_{b}-\Omega_{r}-\Omega_{k})\left[B_{s}+(1-B_{s})a^{-3(1+B)(1+\alpha)}\right]^{\frac{1}{1+\alpha}}\right\}
\end{eqnarray}
where $H$ is the Hubble parameter with its current value $H_{0}=100h\text{km s}^{-1}\text{Mpc}^{-1}$, and $\Omega_{i}$ ($i=b,r,k$) are dimensionless energy parameters of baryon, radiation and effective curvature density respectively. In this paper, we only consider the spatially flat FRW universe.

To study the effects on CMB anisotropic power spectrum, the perturbation evolution equations for MCG would be  studied. We treat MCG as a unified dark fluid which interacts with the rest of matter purely through gravity. With assumption of pure adiabatic contribution to the perturbations, the speed of sound for MCG is 
\begin{equation}
c^{2}_{s}=\frac{\delta p}{\delta \rho}=\frac{\dot p}{\dot \rho}=-\alpha w+(1+\alpha)B,\label{eq:cs2}
\end{equation}
where $w$ is the EoS of MCG in the form of
\begin{equation}
w=B-(1+B)\frac{B_{s}}{B_{s}+(1-B_{s})a^{-3(1+B)(1+\alpha)}}.
\end{equation}
In the synchronous gauge, using the conservation of energy-momentum tensor $T^{\mu}_{\nu;\mu}=0$, one has the perturbation equations of density contrast and velocity divergence for MCG
\begin{eqnarray}
\dot{\delta}_{MCG}&=&-(1+w)(\theta_{MCG}+\frac{\dot{h}}{2})-3\mathcal{H}(c^{2}_{s}-w)\delta_{MCG}\\
\dot{\theta}_{MCG}&=&-\mathcal{H}(1-3c^{2}_{s})\theta_{MCG}+\frac{c^{2}_{s}}{1+w}k^{2}\delta_{MCG}-k^{2}\sigma_{MCG}
\end{eqnarray}
following the notation of Ma and Bertschinger \cite{ref:MB}. For the perturbation theory in gauge ready formalism, please see \cite{ref:Hwang}. The shear perturbation $\sigma_{MCG}=0$ is assumed and the adiabatic initial conditions are adopted in our calculation. 

To keep the perturbation evolution stable, the positivity of sound speed $c_s^2$ is demanded. Actually, to protect the causality, an upper bound $c^2_s\le 1$ is also needed. If one has the assumption of $\alpha\ge 0$ and $0\le B_s\le 1$, one can find easily that $c^2_s$ is non-negative if $B$ respects to 
the inequality $B\geq -\alpha B_s /(1+\alpha B_s )$. However, this assumption is unnatural. Here we shall take $\alpha$ as a free model parameter and remove any unnatural assumption. The detailed methodology for keeping the positivity of sound speed $c_s^2$ will be discribed in the next section.

\section{Constraint method and results}\label{sec:method}

\subsection{Method and data points}

We perform the observational constraints on parameter space by using Markov Chain Monte Carlo (MCMC) method which is contained in a publicly available cosmoMC package \cite{ref:MCMC}, including the CAMB \cite{ref:CAMB} code to calculate the theoretical CMB power spectrum. We modified the code for the MCG as a unified fluid model with its perturbations included. The following $8$-dimensional parameter space  is adopted
\begin{equation}
P\equiv\{\omega_{b},\Theta_{S},\tau, \alpha,B,B_{s},n_{s},\log[10^{10}A_{s}]\}
\end{equation}
where $\omega_{b}=\Omega_{b}h^{2}$ is the physical baryon density, $\Theta_{S}$ (multiplied by $100$) is the ration of the sound horizon and angular diameter distance, $\tau$ is the optical depth, $\alpha$, $B$ and $B_{s}$ are three newly added model parameters related to MCG, $n_{s}$ is scalar spectral index, $A_{s}$ is the amplitude of of the initial power spectrum. Please notice that the current dimensionless energy density of MCG $\Omega_{MCG}$ is a derived parameter in a spatially flat ($k=0$) FRW universe. So, it is not included in the model parameter space $P$. The pivot scale of the initial scalar power spectrum $k_{s0}=0.05\text{Mpc}^{-1}$ is used. We take the following priors to model parameters: $\omega_{b}\in[0.005,0.1]$, $\Theta_{S}\in[0.5,10]$, $\tau\in[0.01,0.8]$, $\alpha\in[-0.1,0.1]$, $B\in[-0.1, 0.1]$, $B_{s}\in[0,1]$, $n_{s}\in[0.5,1.5]$ and $\log[10^{10}A_{s}]\in[2.7, 4]$. In addition, the hard coded prior on the comic age $10\text{Gyr}<t_{0}<\text{20Gyr}$ is imposed. Also, the weak Gaussian prior on the physical baryon density $\omega_{b}=0.022\pm0.002$ \cite{ref:bbn} from big bang nucleosynthesis and new Hubble constant $H_{0}=74.2\pm3.6\text{kms}^{-1}\text{Mpc}^{-1}$ \cite{ref:hubble} are adopted. 

As is seen in Eq. (\ref{eq:cs2}), the expression of $c^2_s$ which contains model parameters $\alpha$, $B$, $B_s$ and scale factor $a$, is complicated. Giving an explicit range of model parameters to keep $c^2_s$ nonnegative is really difficult. Maybe, in some senses, it is impossible. To circumvent the problem, we take the code as a black box, and hard code the condition $c^2_s(a)\ge0$. It means that, in very sampling, the code checks whether the value of $c^2_s$ is negative or not at first. If it is negative, the parameters combination is thrown away and a new one is generated randomly. If it is not negative, the evolutions of background and perturbation for MCG will be calculated. In this way, the output parameters space holds the condition $c^2_s\ge 0$. And we do not worry about its upper bound, because the value of $c^2_s$ is very small and near to zero. So the check condition behaves like a filter. We can check the evolution of $c^2_s$ with respect to scale factor $a$ once the final result is obtained. As is shown in the following section, please see Figure \ref{fig:wdcs}, it really works.

The total likelihood $\mathcal{L} \propto e^{-\chi^{2}/2}$ is calculated to get the distribution, here $\chi^{2}$ is given as
\begin{equation}
\chi^{2}=\chi^{2}_{CMB}+\chi^{2}_{BAO}+\chi^{2}_{SN}.
\end{equation}
The CMB data include temperature and polarization power spectrum from WMAP $7$-year data \cite{ref:lambda} as dynamic constraint. The geometric constraint comes from standard ruler BAO and standard candle SN Ia. For BAO, the values $\{r_{s}(z_{d})/D_{V}(0.2),r_{s}(z_{d})/D_{V}(0.5)\}$ and their inverse covariant matrix \cite{ref:BAO} are used. To use the BAO information, one needs to know the sound horizon at the redshift of drag epoch $z_{d}$. Usually, $z_{d}$ is obtained by using the accurate fitting formula \cite{ref:EH} which is valid if the matter scalings $\rho_{b}\propto a^{-3}$ and $\rho_{c}\propto a^{-3}$ are respected. Obviously, it is not true in our case. So, we find $z_{d}$ numerically from the following integration
\cite{ref:Hamann}
\begin{eqnarray}
\tau(\eta_d)&\equiv& \int_{\eta}^{\eta_0}d\eta'\dot{\tau}_d\nonumber\\
&=&\int_0^{z_d}dz\frac{d\eta}{da}\frac{x_e(z)\sigma_T}{R}=1
\end{eqnarray}   
where $R=3\rho_{b}/4\rho_{\gamma}$, $\sigma_T$ is the Thomson cross-section and $x_e(z)$ is the fraction of free electrons. Then the sound horizon is
\begin{equation}
r_{s}(z_{d})=\int_{0}^{\eta(z_{d})}d\eta c_{s}(1+z).
\end{equation}   
where $c_s=1/\sqrt{3(1+R)}$ is the sound speed. We use the substitution \cite{ref:Hamann}
\begin{equation}
d_z\rightarrow d_z\frac{\hat{r}_s(\tilde{z}_d)}{\hat{r}_s(z_d)}r_s(z_d),
\end{equation}
to obtain unbiased parameter and error estimates, where $d_z=r_s(\tilde{z}_d)/D_V(z)$, $\hat{r}_s$ is evaluated for the fiducial cosmology of Ref. \cite{ref:BAO}, and $\tilde{z}_d$ is obtained by using the fitting formula \cite{ref:EH} for the fiducial cosmology. Here $D_V(z)=[(1+z)^2D^2_Acz/H(z)]^{1/3}$ is the 'volume distance' with the angular diameter distance $D_A$. 
The $557$ Union2 data with systematic errors are also included \cite{ref:Union2}. For the detailed description of SN, please see Refs. \cite{ref:Xu}.

\subsection{Fitting Results and Discussion}

\begin{center}
\begin{figure}[h]
\includegraphics[width=9.5cm]{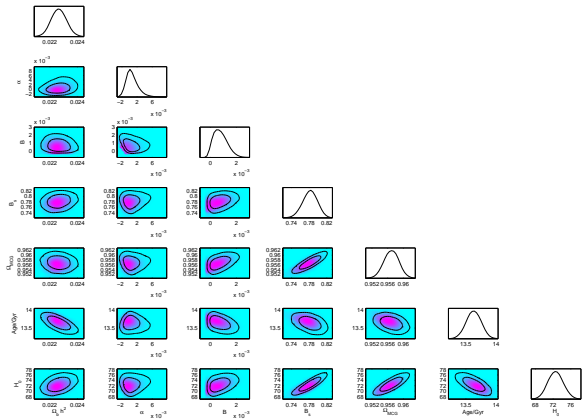}
\caption{The 1D marginalized distribution on individual parameter and 2D contours  with $68\%$ C.L. and $95\%$ C.L. by using CMB+BAO+SN data points.}\label{fig:contour}
\end{figure}
\end{center}

We generate $8$ independent chains in parallel and stop sampling by checking the worst e-values [the
variance(mean)/mean(variance) of 1/2 chains] $R-1$ of the order $0.01$. The calculated results of the model parameters and derived parameters are shown in Table. \ref{tab:results}, where the mean values with $1\sigma$ and $2\sigma$ regions from the combination WMAP+BAO+SN are listed. The minimum $\chi^2$ is $8004.630$ which is smaller than that $\chi^2_{min}=8009.116$ for $\Lambda$CDM model with the same data sets combination. Correspondingly, the contour plots are shown in Figure \ref{fig:contour}.
\begingroup
\begin{table}[tbh]
\begin{center}
\begin{tabular}{cc}
\hline\hline Prameters&Mean with errors\\ \hline
$\Omega_b h^2$ & $    0.0226_{-    0.000536    0.00104}^{+    0.000540+    0.00108}$ \\
$\theta$ & $    1.0490_{-    0.002521-    0.00494}^{+    0.00251+    0.00502}$ \\
$\tau$ & $    0.0883_{-    0.00735-    0.0238}^{+    0.00659+    0.0255}$ \\
$\alpha$ & $    0.000727_{-    0.00140-    0.00234}^{+    0.00142+    0.00391}$ \\
$B$ & $    0.000777_{-    0.000302-    0.000697}^{+    0.000201+    0.000915}$ \\
$B_s$ & $    0.782_{-    0.0162-    0.0329}^{+    0.0163+    0.0307}$ \\
$n_s$ & $    0.987_{-    0.0146-    0.0285}^{+    0.01451+    0.0287}$ \\
$\log[10^{10} A_s]$ & $    3.0844_{-    0.0339-    0.0651}^{+    0.0333+    0.0695}$ \\
$\Omega_{MCG}$ & $    0.957_{-    0.00184-    0.00363}^{+    0.00182+    0.00359}$ \\
$\text{Age/Gyr}$ & $   13.629_{-    0.118-    0.235}^{+    0.120+    0.228}$ \\
$\Omega_b$ & $    0.043_{-    0.00182-    0.00359}^{+    0.00184+    0.00363}$ \\
$z_{re}$ & $   10.524_{-    1.216-    2.400}^{+    1.2152+    2.417}$ \\
$H_0$ & $   72.561_{-    1.690-    3.223}^{+    1.679+    3.361}$ \\
\hline\hline
\end{tabular}
\caption{The mean values of model parameters with $1\sigma$ and $2\sigma$ errors from the combination WMAP+BAO+SN.}\label{tab:results}
\end{center}
\end{table}
\endgroup

\begin{center}
\begin{figure}[tbh]
\includegraphics[width=8.5cm]{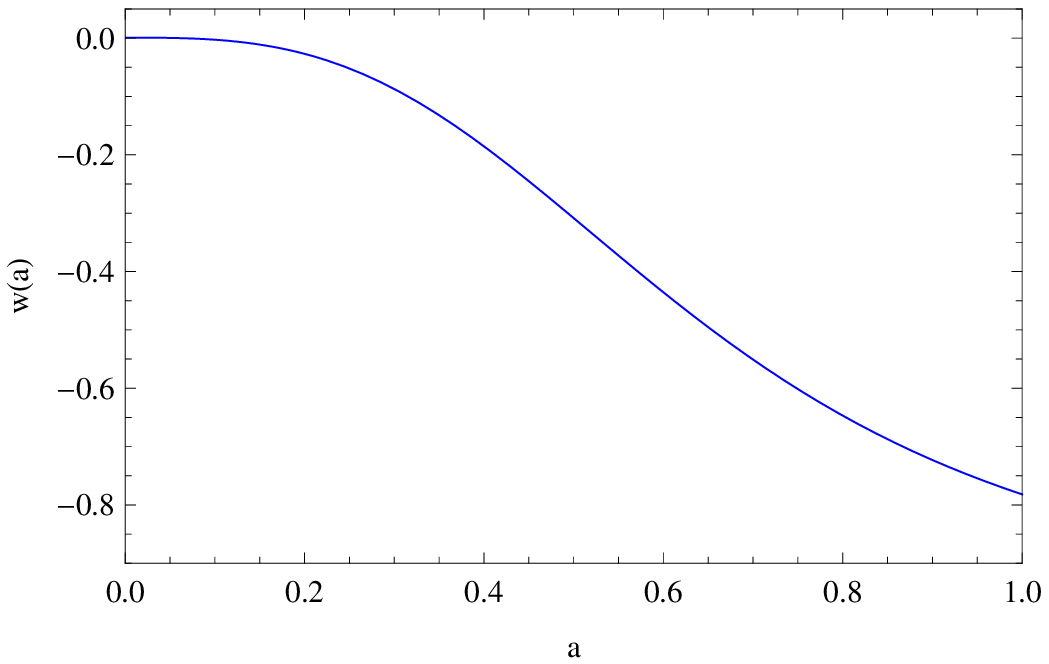}
\includegraphics[width=8.5cm]{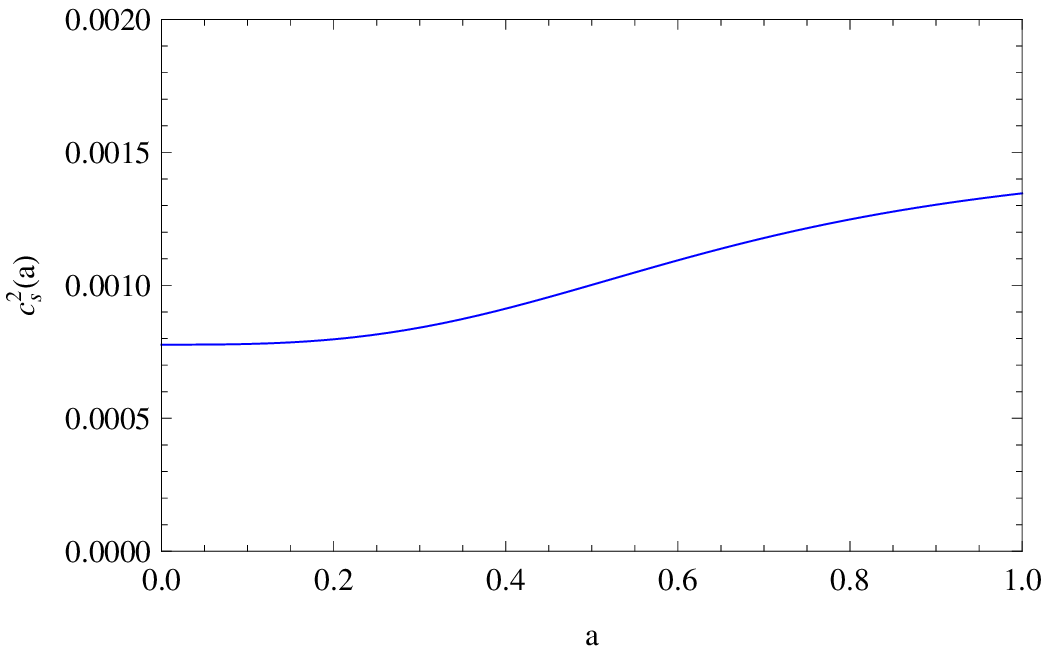}
\caption{The evolutions of EoS $w(a)$ and sound speed $c^2_s(a)$ for MCG with respect to scale factor $a$.}\label{fig:wdcs}
\end{figure}
\end{center}

From the Table \ref{tab:results} and Figure \ref{fig:contour}, one can clearly see that a tight constraint is obtained when the full information of CMB data is included. For the small values of $\alpha$ and $B$, one finds that MCG model is very close to $\Lambda$CDM model. And the current data slightly favor MCG model. Using the mean values of model parameters, we plot the evolution of EoS $w(a)$ and the speed of sound $c^2_s(a)$ for MCG with respect to scale factor $a$ in Figure \ref{fig:wdcs}. From the left panel of Figure \ref{fig:wdcs}, the MCG behaves like cold dark matte at early epoch. The right panel of Figure \ref{fig:wdcs} shows that the value of the sound speed $c^2_s(a)$ of MCG is small positive number and varies with scale factor. The small values of $c^2_s(a)$ make it possible to form large scale structures in our universe. Also, the positivity of $c^2_s(a)$ is really guaranteed by the 'filter' in sampling. 

\begin{center}
\begin{figure}[tbh]
\includegraphics[width=8.5cm]{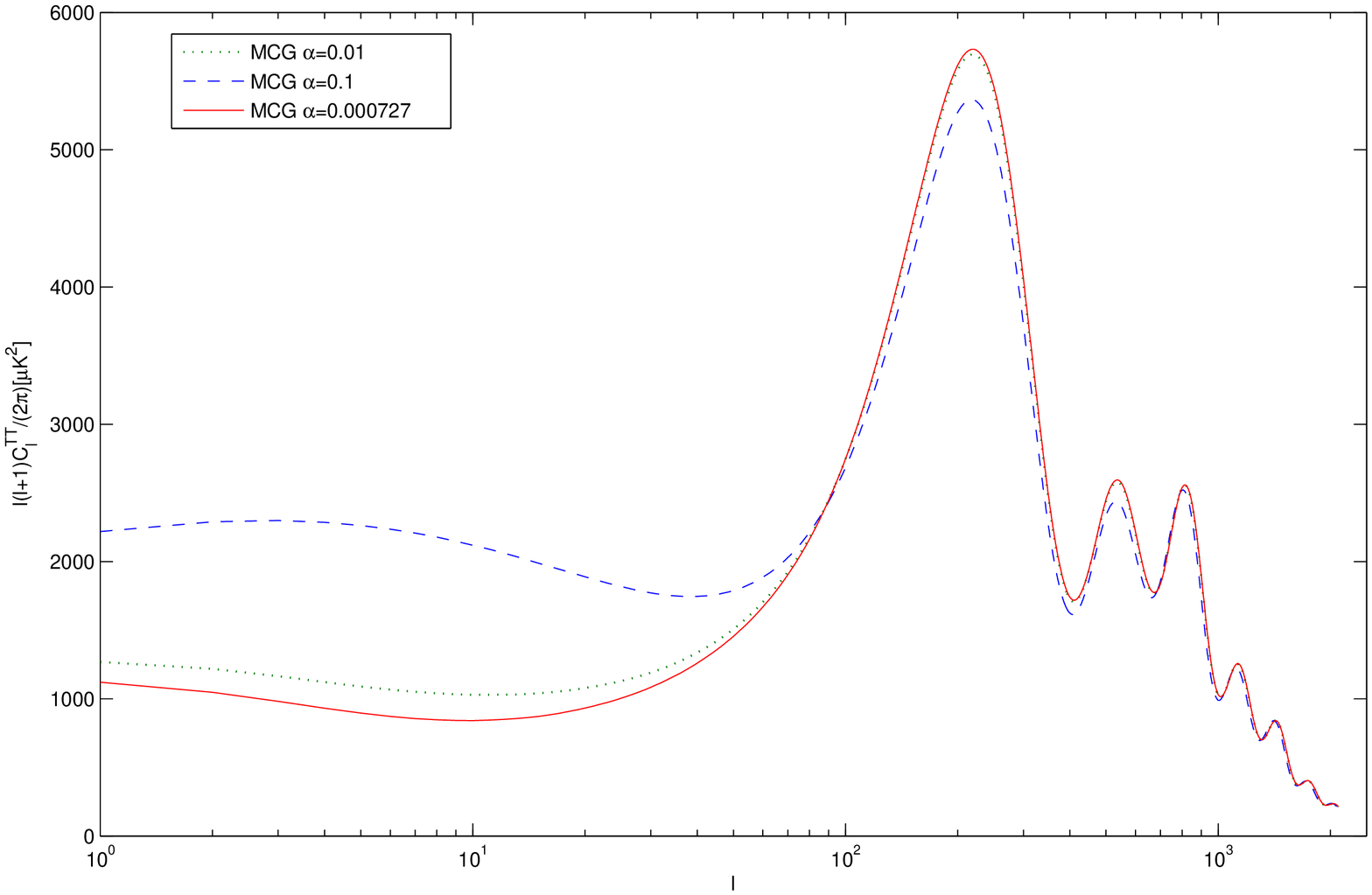}
\includegraphics[width=8.5cm]{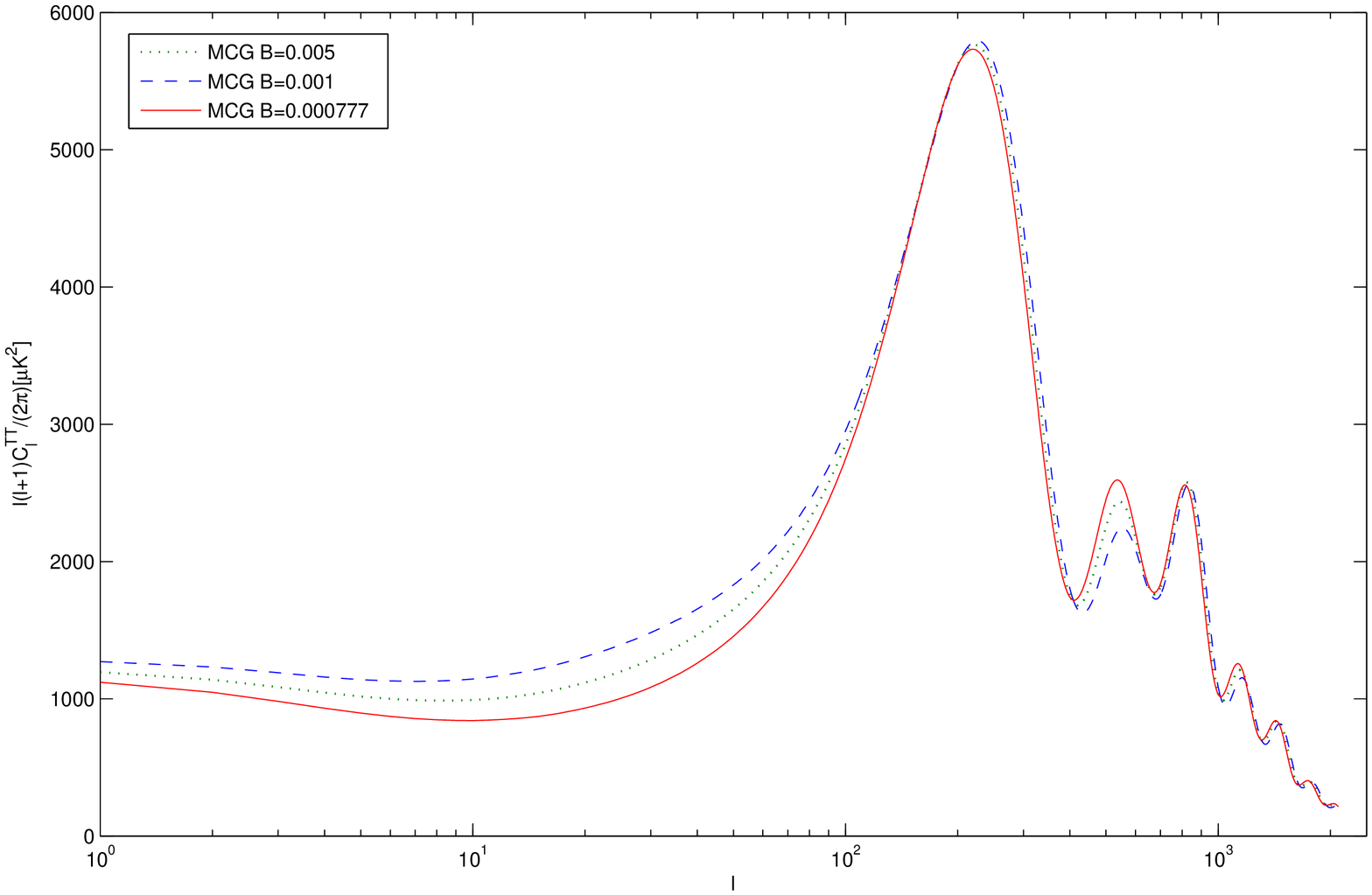}
\includegraphics[width=8.5cm]{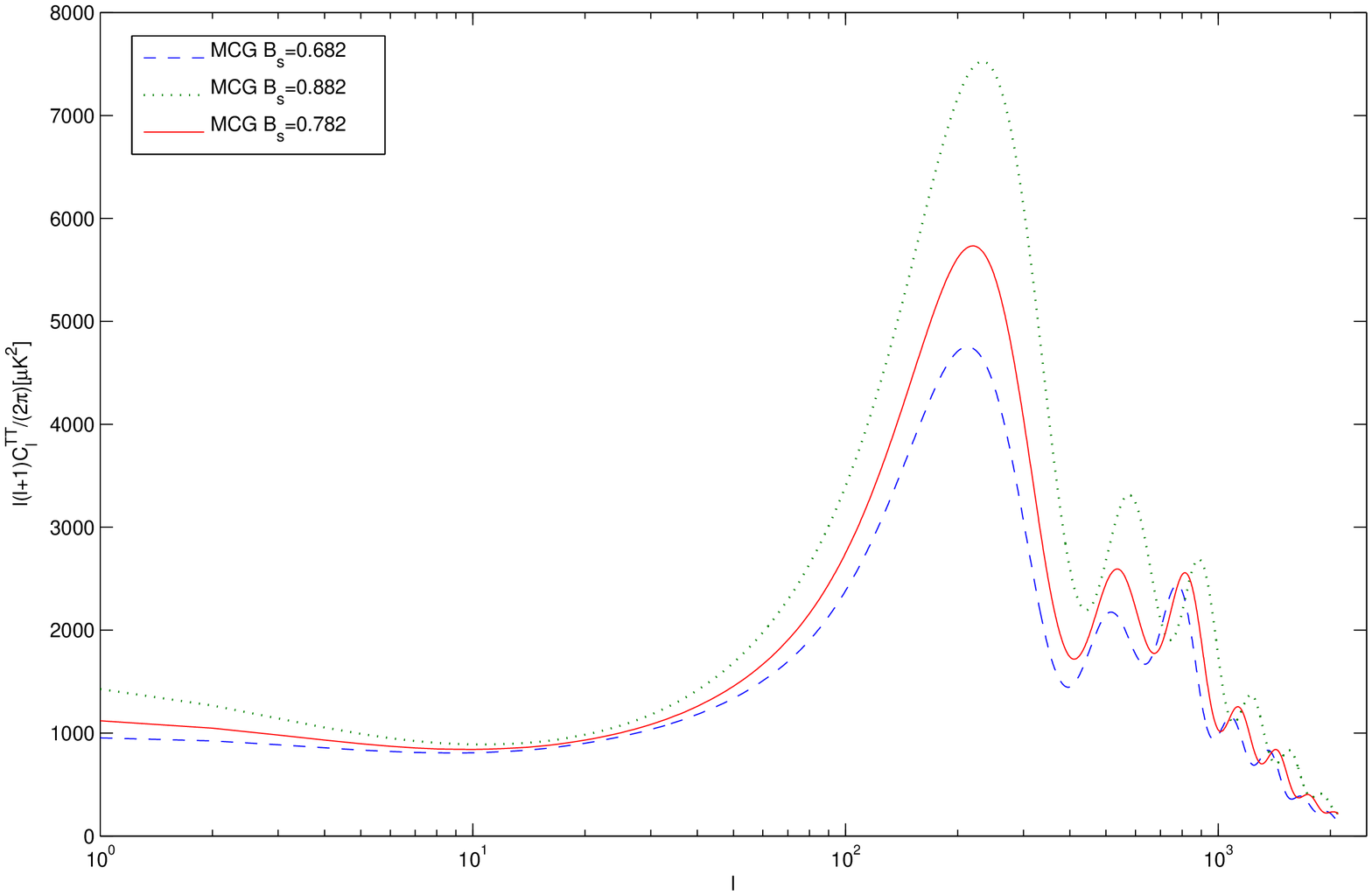}
\includegraphics[width=8.5cm]{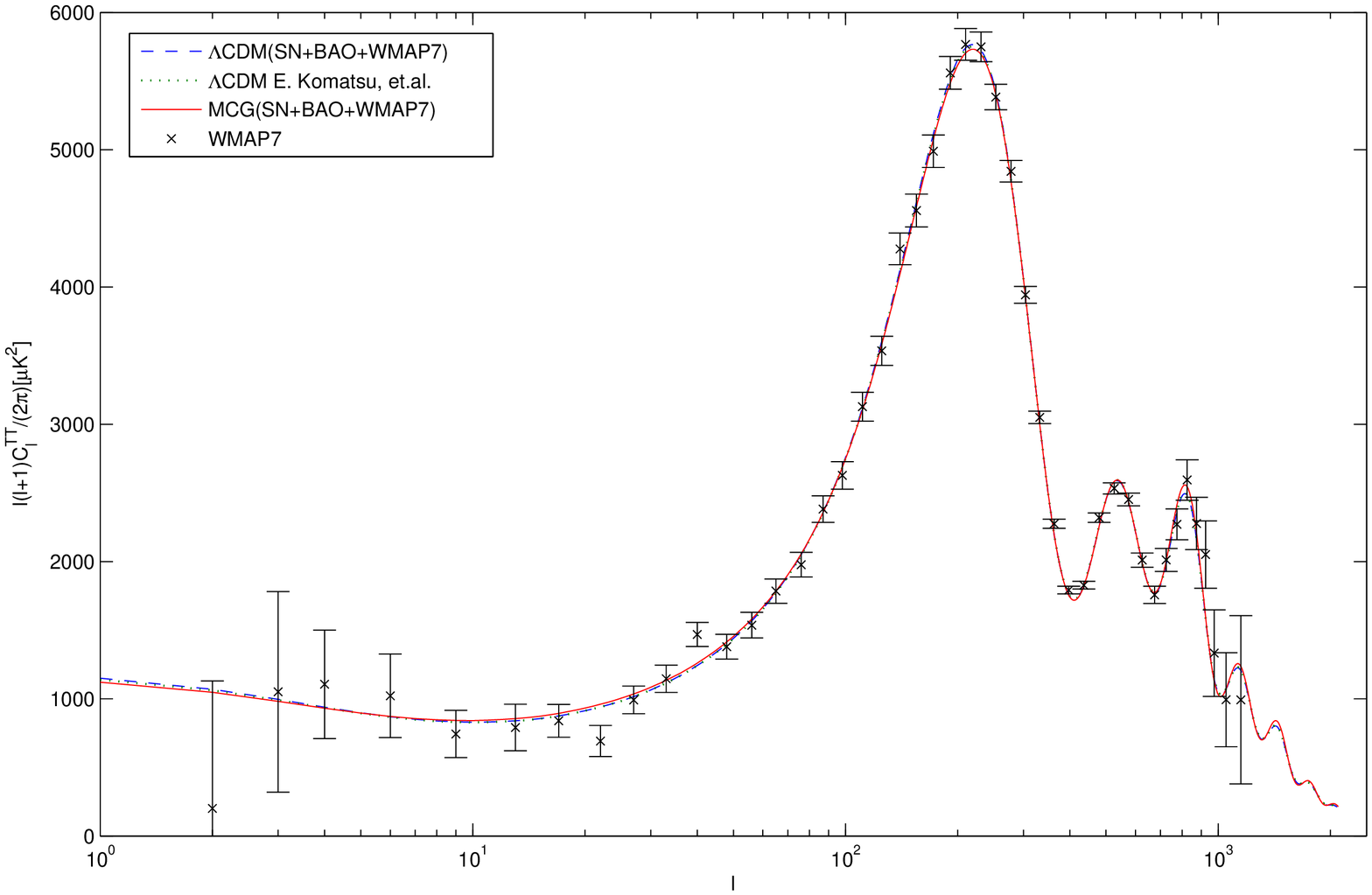}
\caption{The CMB $C^{TT}_l$ power spectrum v.s. multiple moment $l$. The upper left, right and lower left panels show the effects of three model parameters $\alpha$, $B$ and $B_s$ to CMB temperature anisotropic power spectra respectively, in each case the other relevant model parameters are fixed to the mean values as listed in Table \ref{tab:results}. The lower right panel shows CMB power spectra with mean values listed in Table \ref{tab:results} for MCG and $\Lambda$CDM model, where the black dots with error
bars denote the observed data with their corresponding uncertainties from WMAP $7$-year results, the red solid line is for MCG with mean values as shown in Table \ref{tab:results}, the blue dashed line is for $\Lambda$CDM model with mean values for the same data points combination. And the green doted line is for $\Lambda$CDM model with mean values taken from \cite{ref:wmap7} with WMAP+BAO+$H_0$ constraint results.}\label{fig:cls}
\end{figure}
\end{center}

To understand the effects of model parameters to the CMB anisotropic power spectra, we plot the Figure \ref{fig:cls}, where one of three model parameters $\alpha$, $B$ and $B_s$ varies in the first three panels, and the other relevant parameters are fixed to their mean values as listed in Table \ref{tab:results}. The upper left and right panels of Figure \ref{fig:cls} show the effect of parameter $\alpha$ and $B$ to CMB power spectra respectively. The model parameters $\alpha$ and $B$ modify the power law of the energy density of MCG, then they make the gravity potential evolution at late epoch of the universe. As results, one can see Integrated SachsÐWolfe (ISW) effect on the large scale as shown in the upper left and right panels of Figure \ref{fig:cls}. In the early epoch, MCG behaves like cold dark matter with almost zero EoS and speed of sound $c^2_s$ as shown in Figure \ref{fig:wdcs}, therefore the variation of the values of $\alpha$ and $B$ will change the ratio of energy densities of the effective cold dark matter and baryons. One can read the corresponding effects from the variation of the first and the second peaks of CMB power spectra. The parameter $B_s$ is related with the dimensionless density parameter of effective cold dark matter $\Omega_{c0}$. Decreasing the values of $B_s$, which is equivalent to increase the value of effective dimensionless energy density of cold dark matter, will make the equality of matter and radiation earlier, therefore the sound horizon is decreased. As a result, the first peak is depressed. The lower right panel shows 
CMB power spectra with mean values listed in Table \ref{tab:results} for MCG and $\Lambda$CDM model, where the black dots with error
bars denote the observed data with their corresponding uncertainties from WMAP $7$-year results, the red solid line is for MCG with mean values as shown in Table \ref{tab:results}, the blue dashed line is for $\Lambda$CDM model with mean values for the same data points combination. And the green doted line is for $\Lambda$CDM model with mean values taken from \cite{ref:wmap7} with WMAP+BAO+$H_0$ constraint results. One can see that MCG can match observational data points and $\Lambda$CDM model well. This is the evidence of dark degeneracy. 

\section{Summary} \label{ref:conclusion} 

In summary, we perform a global fitting on MCG model, which is treated as a unified dark matter and dark energy model, by using MCMC method with the combination of the full CMB, BAO and  SN Ia data points. As a contrast to the reports in the literatures, we take MCG as an entire energy component and without any decomposition. Tight constraint is obtained as shown in Table \ref{tab:results} and Figure \ref{fig:contour}. The MCG model can match observational data points and $\Lambda$CDM model well. This is the dark degeneracy. For the small values of model parameter $\alpha$ and $B$, one can conclude that MCG model is very close to $\Lambda$CDM model. And the current data favor MCG model slightly. 

\section{Acknowledgements} We thank an anonymous referee for helpful improvement of this paper. L. Xu's work is supported by the Fundamental Research Funds for
the Central Universities (DUT10LK31) and (DUT11LK39). H. Noh's work is supported by Mid-career Research Program through National
Research Foundation funded by the MEST (No. 2010-0000302).

\end{document}